\documentclass[dvips]{elsart4}

\usepackage{graphicx}
\usepackage{amssymb}

\begin{document}
\begin{frontmatter}

%----------------------------------------------------------------------
% Specify destination and version number of the manuscript

%----------------------------------------------------------------------
% Title of manuscript

\title{Arsenic uptake by gypsum and calcite:\\
Modeling and probing by neutron and x-ray scattering}

%----------------------------------------------------------------------
% List of authors
%
% List each author using a separate \author{} command
%
% If there is more than one author address, add a label to each author
% of the form \author[label]{name}.  This label should be identical to
% the corresponding label provided with the \address command.
%
% e.g. if there are three authors from two institutions in USA and 
% France, you can link them to their respective addresses, using
%
% \author[US]{John Doe}
% \author[US,FR]{Jane Doe}
% \author[FR]{Jean Dupont}
% \address[US]{University of Life, Somewhere, USA}
% \address[FR]{Universite de la Vie, Quelque Part, France}
%
% N.B. Unlike the document class used for abstract submissions, it is
% possible to have the author associated with more than one address,
% as shown in the example above.
%

\author[ILL]{A. Fern\'andez-Mart\'{\i}nez}
\author[GIR]{G. Rom\'an-Ross}
\author[ILL]{G. J. Cuello\corauthref{1}}
\author[CSIC]{X. Turrillas}
\author[LGIT]{L. Charlet}
\author[ILL]{M. R. Johnson}
\author[GILDA]{F. Bardelli}

%----------------------------------------------------------------------
% List of addresses
%
% If there is more than one address, list each using a separate 
% \address command using a label to link it to the respective author
% as described above
 
\address[ILL]{Institut Laue-Langevin, B.P. 156, 38042 - Grenoble Cedex 9 (France)}
\address[GIR]{Universidad de Girona, Campus de Montilivi, 17071 - Girona (Spain)}
\address[CSIC]{ICC Eduardo Torroja, CSIC, Serrano Galvache s/n, 28033 - Madrid (Spain)}
\address[LGIT]{LGIT-IRIGM, Univ. de Grenoble I, B.P. 53, 38041 - Grenoble Cedex 9 (France)}
\address[GILDA]{INFM-GILDA c/o ESRF, B.P. 220, 38043 - Grenoble Cedex 9 (France)}

%----------------------------------------------------------------------
% Title page footnotes
%
% If you need to add qualifying information to any of the authors, 
% use the \thanksref{} command within the \author command.  The 
% argument is the label of a corresponding \thanks[label]{text}
% command which contains the footnote text
%
% e.g. you can acknowledge a funding authority for John Doe, using
%
% \author{John Doe\thanksref{ABC}}
% \thanks[ABC]{This work was supported by Institute of Unphysical 
%    Phenomena under contract no. ABC-123}
%

%\thanks[]{}

%----------------------------------------------------------------------
% Contact Information
%
% Add the complete postal address, telephone number, fax number, and
% email address of the corresponding author as a special footnote using
% the \corauth[]{} command.  This works in a similar way to the \thanks 
% command.  Add the \corauthref{} command within the \author command.
% The argument is the label of a corresponding \corauth[label]{text}
% command which contains the contact information.  Prefix the text with
% Corresponding Author:
%
% e.g. if the contact author is John Doe,
%
% \author{John Doe\corauthref{1}}
% \corauth[1]{Corresponding Author: University of Life, 123 Some St.,
%    Somewhere, MI 12345, USA.  Phone: (555) 555-5555 
%    Fax: (555) 555-7777, Email: JDoe@uol.edu}
%

\corauth[1]{Corresponding Author: Institut Laue-Langevin, B.P. 156,
Grenoble Cedex 9, 38042 France.  Phone: (0033) 4 76 20 76 97 
Fax: (0033) 4 76 20 76 48, Email: cuello@ill.fr}

%----------------------------------------------------------------------
% Text of abstract

\begin{abstract}

Uptaking of contaminants by solid phases is relevant to many issues in 
environmental science as this 
process can remove them from solutions and retard their transport into the 
hydrosphere. Here we report on two structural studies performed on 
As-doped gypsum 
(CaSO$_4$ 2H$_2$O) and calcite (CaCO$_3$), using neutron (D20-ILL) 
and x-ray (ID11-ESRF) 
diffraction data and EXAFS (BM8-ESRF). The aim of this study is to determine 
whether As gets 
into the bulk of gypsum and calcite structures or is simply adsorbed on the 
surface. Different mechanisms of substitution are used as hypotheses. 
The combined Rietveld analysis of 
neutron and x-ray diffraction data shows an expansion of the unit cell 
volume proportional to the As concentration 
within the samples. DFT-based simulations confirm the increase of the unit 
cell volume proportional to the 
amount of carbonate or sulphate groups substituted. 
Interpolation of the experimental Rietveld data allows us to distinguish 
As substituted within the structure from that adsorbed on 
the surface of both minerals. Results obtained by EXAFS analysis from 
calcite samples show good 
agreement with the 
hypothesis of replacement of As into the C crystallographic site.

\end{abstract}

%----------------------------------------------------------------------
% Manuscript keywords
%
% Please give two or three keywords in the form: keyword \sep keyword
% e.g. NMR \sep superconductivity
%
% NB The syntax is different from the abstract document class


%----------------------------------------------------------------------
% End of front page

\end{frontmatter}

%----------------------------------------------------------------------
% Manuscript text
%
% Fill in the following space with the manuscript text.
%
% A number of LaTeX commands may be invoked in this space, e.g.
%
% \section{} : to insert a new section title
% \label{}   : to label the numbered section for use in \ref{}
% \cite{}    : to add a reference using the label in \bibitem{}
% 
% A number of LaTeX environments may be used, e.g. 
% \begin{equation}
%     An equation inserted here will be automatically numbered
% \end{equation}  
%
% Please refer to other LaTeX documentation for help on using these
% environments.

\section{Introduction}
Arsenic is recognized as a dangerous pollutant of the environment \cite{Nriagu}.
Arsenic present in groundwaters may be trapped in the solid phase of minerals like calcite or gypsum, either by adsorption or by co-precipitation. %
%Uptake of contaminants in solid phases can remove them from solution retarding their transport.
When a contaminant is incorporated in the bulk rather than simply adsorbed at the surface, it is less available and it can be considered ``immobilized'' in the environment at least until the host phase dissolution. 
The aim of this study is to elucidate whether the incorporation of As(III) and As(V) into the bulk of calcite and gypsum, respectively, occurs or not, and to what extent.

Surface-sensitive X-ray Standing Wave (XSW) studies by Cheng \textit{et al}. \cite{Cheng} show that the As atom replaces the C atom in the carbonate molecules of calcite. The geometry of the carbonate group is not preserved, showing a displacement of the As atom of 0.76 \AA$\:$ in the $[001]$ direction. Density Functional Theory (DFT) based simulations have proved that this replacement drives to a similar displacement of the As atom when C atoms are replaced by As atoms in the bulk of calcite (see below). This fact leads us to keep the same hypothesis of arsenite/carbonate replacement in our study of As incorporation into the bulk.
 
The stability diagram of aqueous solutions of H$_n$AsO$_4^{n-3}$ shows that the arsenate (AsO$_4^{3-}$) is the most stable specie under oxidizing conditions. This anion is a tetrahedron with the As atom at the centre, surrounded by four O atoms. The fact that both, arsenate and sulphate groups have the same geometry supports the hypothesis of a possible replacement of sulphate by arsenate groups when gypsum is precipitated in presence of As(V). The charge is compensated by bonding to an extra proton.

\section{Materials and methods}
In order to test the possible mechanisms for As immobilization by calcite and gypsum, samples of both minerals were synthesized in the presence of As(III) and As(V), respectively. Calcite precipitation was conducted at pH = 7.5 by addition of CaCl$_2$ and Na$_2$CO$_3$ solutions. Gypsum was precipitated from 
Na$_2$SO$_4$ and CaCl$_2$ solutions at three different pH values: 4, 7.5 and 9. Arsenic concentrations incorporated in the solids range between 30 mM/kg and 1200 mM/kg for calcite and 100 mM/kg to 1000 mM/kg for gypsum.

Powder samples were analysed by neutron diffraction at the high flux powder diffractometer D20 (ILL).
Experiments were carried out using a Cu(200) monochromator which gives a wavelength of $\lambda = 1.30$ \AA$\:$ and at ambient conditions of pressure and temperature \cite{calcite}. 
Diffraction patterns were taken for the samples in their container and for the empty cell, in the range of 10$^\circ$ to 130$^\circ$. 
Also powder diffraction experiments were performed with x-ray at ID11 (ESRF), reproducing
the same experimental conditions but using a wavelength of $\lambda = 0.5$ \AA. Both %
%($0.51819 \rm\:\AA$ after the refinement of a standard of Si). 
%The measurement range was 2$^\circ$ - 60$^\circ$ (in $2\theta$) with a step of 0.004$^\circ$. 
diffraction data sets were analysed using Fullprof \cite{Fullprof}.

Geometrical optimisations of the unit cell and the supercells of pure and As-doped calcite and gypsum were done with the Vienna Ab-initio Simulation Package (VASP) \cite{Kresse1}. The PBE functional and PAW pseudopotentials were used. The goal was to reproduce the expansion of the unit cell produced by the incorporation of As atoms into the structure of both minerals. Unit cells of pure calcite and gypsum obtained from Rietveld refinements were used as starting point for all the models. Geometrical optimizations of single unit cells and of supercells of gypsum were done replacing the sulphate molecules SO$_4^{2-}$ by arsenates AsO$_4^{3-}$. 
Similar simulations were performed with supercells of calcite replacing C by As atoms \cite{calcite}.

EXAFS data were collected on a diffraction and absorption beamline (GILDA-BM8) at the ESRF of Grenoble and extracted using standard procedures \cite{Lee}. The theoretical photoelectron paths were generated using the FEFF8 \cite{Ankudinov} code and the fit performed using the MINUIT library from CERN \cite{James}. A monochromator of Si(311) was used to set the incident energy at the K-edge of As (11867 eV).

\section{Results and discussion}

Our diffraction data show an expansion of the unit cell due to As incorporation into calcite crystallites (Fig. 1). By modelling, As concentration in the bulk of the samples can be extrapolated by comparison of the relative volume changes between the experimental and simulated data. 
Calcite unit cell and two supercells (2x2x1 and 3x2x1) were geometrically optimised replacing one and two units of AsO$_3^{3-}$ (150 mM/kg and 290 mM/kg) by CO$_3^{2-}$ units. 
The simulations showed a volume expansion linearly dependent on the replacement of As, 
as expected by Vegard's law (Fig. 2). 
This augmentation is due to the lattice expansion along the $a$ axis as As concentration in 
solids increases. 
The observed As ion displacement of 0.57 \AA$\:$ over the O base along the $(0001)$ is compatible with  Cheng's results \cite{Cheng}. The lower value of this displacement (0.57 \AA$\,$ {\it vs}. 0.76 \AA) can be due to the fact that atoms near the surface are less attached to the solid and can move more freely.

\begin{figure}
    \centering
    \includegraphics[width=0.45\textwidth]{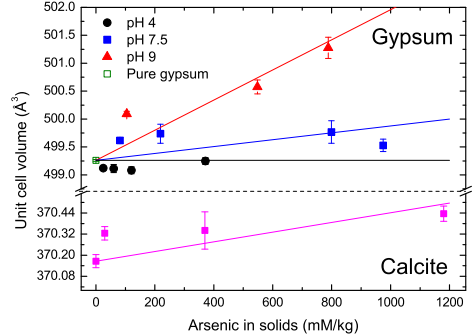}
    \caption{Experimental unit cell volume of As-doped calcite and gypsum samples in function of As concentration.} 
\end{figure}  

The experimental value of the unit cell volume was interpolated using the linear fit of the simulated  volume expansion, giving values of 9, 10 and 16 mM/kg of As in calcite for the three synthesised samples.
Simulations of 3x2x1 supercells with one CO$_3^{2-}$ unit replaced by one AsO$_3^{3-}$ unit were done to check whether the replacement is more likely in sites at the same or at different crystallographic planes. The higher enthalpy of formation ($\Delta H = 211.36$ meV) for the calcite structure with two As atoms lying on the same plane shows that replacement is more likely to happen in different planes, leading to a more stable structure.

We found from EXAFS data analysis a nearest neighbour distance of $\approx 1.77(3)$ \AA$\:$ corresponding to the As-O bond distance. This value lies in between the reported one \cite{Loewenschuss} for the arsenite molecule ($d_{\rm As-O}^{\rm L} = 1.84$ \AA) and that obtained from the simulations 
($d_{\rm As-O}^{\rm sims} = 1.75$ \AA). The coordination number was kept fixed to its theoretical value 
($N = 3$) in order to reduce the correlation between free parameters in the fitting procedure. This result supports the hypothesis of the incorporation of the As atoms into the C crystallographic sites.

\begin{figure}
    
    \includegraphics[width=0.45\textwidth]{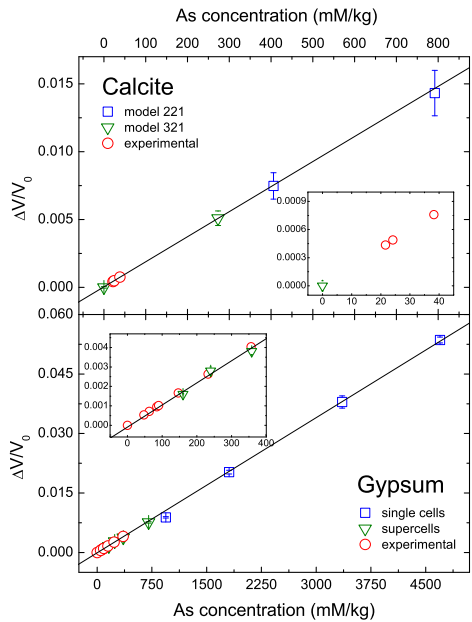}
    \caption{Simulated relative expansion of unit cell volume of gypsum and calcite. The insets show a detail of the interpolation of the experimental data using the simulated expansion.}
\end{figure}  

Figure 1 shows the gypsum unit cell volume obtained from combined refinements of neutron and x-ray data. The expansion of the unit cell is proportional to the As concentration in solids and strongly dependent on the pH value: the biggest expansion is found in samples synthesized at pH 9. This result is in good agreement with the hypothesis of replacement of sulphate (SO$_{4}^{2-}$) by arsenate groups (AsO$_{4}^{3-}$). This replacement is more likely at higher pH values, according to available thermodynamical data regarding speciation of As \cite{Pourbaix}. The expansion of the unit cell parameters is due to the different lengths for As-O ($\approx 1.70$ \AA) \cite{Kolitsch} and S-O ($\approx 1.50$ \AA) \cite{Loureiro} bonds.

Simulations show an increasing of the unit cell volume proportional to the number of atoms of S replaced by As (Fig. 2). Four single cells (with 940, 1809, 3357 and 4696 mM/kg of As) and four supercells were simulated: two 2x1x2 supercells with one and two As atoms, a 2x1x3 and a 3x1x3 with one As atom each, giving As concentrations of 358, 705, 240 and 160 mM/kg, respectively.
%two 2x1x2 supercells with one (358 mM/kg) and two (705 mM/kg) As atoms, and a 2x1x3 (240 mM/kg) and 3x1x3 %(160 mM/kg) with one As atom each. 
The simulations allow us to extrapolate the As concentration in the bulk of the samples by comparing the relative volume variations between the experimental and simulated data (Table 1).

\begin{table}[h]
\centering
\begin{tabular}{|c|c|c|c|}
\hline
Initial \%As & pH 4 & pH 7 & pH 9 \\ \hline
 (M) & \%As (mM/kg)& \%As (mM/kg)& \%As (mM/kg)\\
\hline
0.01 & 0 & 62 & 145 \\
0.04 & 0 & 84 & 232 \\
0.06 & 0 & 46 & 355 \\
0.09 & 0 & 89 & - \\ \hline
\end{tabular}
\caption{Extrapolated values of the As concentration in samples of gypsum. 
The result for pH 4 is due to the near zero relative expansion of the unit cell.}
\end{table}

Our results support the hypothesis of As immobilisation by incorporation into the bulk of these minerals.
This improves the knowledge on the long term stability of contaminated sludges and it has important consequences for site remediation actions. This work is an example of a direct link between fundamental research and environmental issues. The understanding of the As compounds behaviour in sedimentary environments is essential
to estimate and predict possible consequences of forecast or accidental events.

%----------------------------------------------------------------------
% Reference section
%
% List each reference with a separate \bibitem{} command.  The
% argument contains the label that is used in the \cite{} command
% in the main text
%
% e.g.
%
%    This follows our pioneering work on TdB2\cite{TdB2}.
%
% \bibitem{TdB2}
% J. Doe, J. Doe, and J. Dupont, J. Irrep. Res. 10 (2000) 1000.

%----------------------------------------------------------------------
% Figures and Tables
%
% Insert figures and tables at the end of the document unless you
% are familiar with the LaTeX positional options.
%
% \begin{figure}
%     \centering
%     \includegraphics{filename.eps}
%     \caption{Insert figure caption here} 
% \end{figure}  
%
% \begin{table}
%     \centering
%     \begin{tabular}
%     Insert table here
%     \end{tabular}
%     \caption{Insert table caption here}
% \end{table}  
%
% Please refer to other LaTeX documentation for help on using these
% environments.

%----------------------------------------------------------------------
% Terminate document

\end{document}